\documentclass{article}
\usepackage{amsmath}
\usepackage{hhline}
\usepackage{graphicx}
\usepackage[T2A]{fontenc}
\usepackage[cp1251]{inputenc}
\usepackage[russian]{babel}

\begin{document}          \parindent=7mm

\def\ds{\displaystyle}    \def\ss{\scriptstyle}
\def\hh{\hskip 1pt}       \def\hs{\hskip 2pt}   \def\h{\hskip 0.2mm}
\def\pr{\prime}
\newcommand{\blR}{\hbox{\ss\bf R}}
\newcommand{\fbr}{\hbox{\footnotesize\bf r}}
\newcommand{\fbk}{\hbox{\footnotesize\bf k}}
\newcommand{\Tr}{\mathop{\rm T\h r}\nolimits}
\newcommand{\fbR}{\hbox{\footnotesize\bf R}}

\setcounter{page}{1}

\centerline{\large\bf QUANTUM MARKOVIAN KINETIC EQUATION }

\vskip 1 mm

\centerline{\large\bf FOR HARMONIC OSCILLATOR } \vskip 6 mm

\centerline{\large Boris V. Bondarev }\vskip 5mm

\centerline{\it Moscow Aviation Institute, Volokolamsk road, 4,
125871, Moscow, Russia }\vskip 1mm

\centerline{E\hh -mail: bondarev.b@mail.ru }

\vskip 5mm

\begin{quote} \hskip 7mm \small Specific nonequilibrium states of
the quantum harmonic oscillator described by the Lindblad equation
have been hereby suggested. This equation makes it possible to
determine time-varying effects produced by statistical operator or
statistical matrix. Thus, respective representation-varied
equilibrium statistical matrixes have been found. Specific mean
value equations have been found and their equilibrium solutions have
been obtained. \end{quote} \vskip 3mm


\vskip 5 mm \centerline{\bf 1. Lindblad equation } \vskip 2 mm

\par Statistical operator $\hat\varrho$ or statistical matrix is
basically applied as the quantum mechanics tool, any information of
the nonequilibrium process proceeding within the tested system may
be gained from {\linebreak } [1 - 5]. When the process concerned
proceeds within the system which fails interacting with its
environment, statistical operator $\hat\varrho$ will satisfy
Liouville-von Neumann equation as follows: $$
i\hh\hbar\hs\dot{\hat\varrho}= \bigl[\hh\widehat
H,\hh\hat\varrho\hs\bigr]\hs . \eqno (1.1) $$ With provision for the
fact that the system interacts with any environment, a new equation
shall be produced [4 - 16]. Lindblad is the first one who offered
the equation describing interaction of the system with a thermostat
[10]. This work is devoted to Markovian equation, which hereby
describes nonequilibrium quantum harmonic oscillator performance

\par We will write the kinetic equation for a quantum harmonic
oscillator as follows: $$ i\hs\hbar\hskip 1.1mm\dot{\hskip
-0.5mm\hat\varrho}= \bigl[\hh\widehat
H,\hh\hat\varrho\hs\bigr]+i\hs\hbar\hs A\hs\Bigl(\hh \bigl[\hs\hat
a\hs\hat\varrho\hh ,\hs\hat a^{\hh +}\bigl]+ \bigl[\hs\hat a\hh
,\hs\hat\varrho\hs\hh\hat a^{\hh +}\bigl]\hs\Bigr)+i\hs\hbar\hs
B\hs\Bigl(\hh \bigl[\hs\hat a^{\hh +}\hs\hat\varrho\hh ,\hs\hat
a\bigl]+ \bigl[\hs\hat a^{\hh +},\hs \hat\varrho\hs\hh\hat
a\bigl]\hs\Bigr)\hs , \eqno (1.2) $$ where $$ \widehat
H=\hbar\hs\omega\Biggl(\hat a^{\hh +}\hs\hat a+\frac{1}{2}
\Biggr)\hs , \eqno (1.3) $$ $A$ and B are constants. Operator $\hat
a$ is formulated as follows: $$ \hat a=\frac{1}{\sqrt{\hh
2\hs\hbar\hs\omega\hh }}\hs\Biggl( \hh\frac{i\hs\hat p }{\sqrt{\hh
m}}+\sqrt{\hh\kappa}\hs\hat x\Biggr)\hs . \eqno (1.4) $$ Equation
(1.2) is very precise to describe time-varying state of the
thermostat-interacted quantum harmonic oscillator and its
equilibrium state.

\vskip 5mm

\centerline{\bf 2. Energy representation } \vskip 2mm

\par Now, we will define the wave functions describing specific
energy state $\varphi_n(x)$. The very functions satisfy the equation
as follows: $$ \widehat H\hs\varphi_n(x)=E_n\hh\varphi_n(x)\hs ,
\eqno (2.1) $$ where $$ E_n=\hbar\hs\omega\hs\biggl(n+\frac{1}{\hh
2\hh}\biggr)\hs , \hskip 10mm n=0, 1, 2, ... \eqno (2.2) $$

\par As referred to energy representation, the matrix elements of
statistical operator $\hat\varrho$ will be formulated by the
equation as follows: $$ \varrho_{\h
nn^{\pr}}=\int\varphi^*_n(x)\hs\hat\varrho\hs \varphi_{n^\pr}(x)\hs
dx\hs . \eqno (2.3) $$ Wave functions satisfy the following
equations $$ \hat a\hs\varphi_n=\sqrt{\hh n\hh}\hs\varphi_{n-1}\hs ,
\hskip 15mm \hat a^+\hh\varphi_n=\sqrt{\hh
n+1\hh}\hs\varphi_{n+1}\hs . \eqno (2.4) $$ With provision for the
above formulas the following matrix-formed equation (1.2) is
derived: $$ \dot\varrho_{\h nn^{\pr}}=-\hs i\hs\omega\hs(n-n^\pr)\hs
\varrho_{\h nn^{\pr}}+A\hs\Bigl(2\hs\sqrt{(n+1)(n^\pr+1)}\hs
\varrho_{\h{n+1,\hh n^{\pr}+1}}-(n+n^\pr)\hs\varrho_{\h nn^{\pr}}
\Bigr)\hs + \hs $$ $$ +\hs B\hs\Bigl(\hh 2\hs\sqrt{\hh n\hh
n^\pr\hh}\hs \varrho_{\h n-1,\hs
n^{\pr}-1}-(n+n^\pr+2)\hs\varrho_{\h nn^{\pr}}\Bigr) \hs . \eqno
(2.5) $$

\par Now, we will write the equation for diagonal elements of
statistical matrix $\varrho_{\h nn}=W_n$, where $W_n$ is the
probability referred to oscillator state $\varphi_n$. The equation
produced has the form as follows: $$ \dot W_n=2\hs
A\hs\bigl((n+1)\hs W_{n+1}-n\hs W_n\bigr)+ 2\hs B\hs\bigl(n\hs
W_{n-1}-(n+1)\hs W_n\bigr)\hs . \eqno (2.6) $$ This kinetic equation
describes particular harmonic oscillator state transitions. In this
case, there may be gained coefficients $A$ and $B$ as follows $$
A=\frac{1}{\hh 2\hh}\hs P\hs e^{\hh\frac{1}{2}\hh\beta\hh\hbar\hh
\omega}\hs , \hskip 15mm B=\frac{1}{\hh 2\hh}\hs P\hs
e^{\hh-\hh\frac{1}{2}\hh\beta\hh\hbar\hh
 \omega}\hs , \eqno (2.7) $$
where $P$ is probability of transition per unit time; $\beta=1/{\hh
k\hs T\hh}$ is reciprocal temperature.

\par Equation (2.6) has specific oscillator state equilibrium
distribution, which satisfies the following equation $$
A\hs\bigl((n+1)\hs W_{n+1}-n\hs W_n\bigr)+ B\hs\bigl(n\hs
W_{n-1}-(n+1)\hs W_n\bigr)=0 \hs . \eqno (2.8) $$ This equation is
solved by the method as follows $$ W_n=(1-q)\hs q^n \eqno (2.9) $$
under the following condition $$ q=\frac{\hh B\hh}{A}=\exp\hh(\hh
-\hs\beta\hs\hbar\hs\omega)\hs . \eqno (2.10) $$

\vskip 5mm

\centerline{\bf 3. Mean value of coordinate } \vskip 2mm

\par Mean value $\overline b$ assigned by operator $\hat b$ is
defined as $$ \overline b=Tr\hh (\hh\hat b\hs\hat\varrho\hh )\hs .
\eqno (3.1) $$

\par For gaining mean value $\overline a$ the respective equation
may be derived from formula (1.2). Using the equality of: $$ \hat
a\hh\hat a^{\hh +}-\hat a^{\hh +}\hh\hat a=1\hs , \eqno (3.2) $$ we
will get the equation as follows: $$ \dot{\overline a}=(\hh -\hs
i\hs\omega-A+B\hh)\hs\overline a\hs . \eqno (3.3) $$

\par Now, we can find the derivatives from mean values $\overline x$
and $\overline p$. By applying formula (1.4) we will get: $$
\frac{i\hs\dot{\overline p}}{\sqrt{\hh
m}}+\sqrt{\hh\kappa}\hs\dot{\overline x}=(\hh -\hs
i\hs\omega-A+B\hh)\hs\Biggl( \hh\frac{i\hs{\overline p}}{\sqrt{\hh
m}}+\sqrt{\hh\kappa}\hs{\overline x}\Biggr)\hs . $$ Then, we will
try to equate both the real and imaginary parts of this equation: $$
\left.\begin{array}{rcl} \dot{\overline x}&=&-\hs (A-B)\hs{\overline
x}+\overline p/m\hs , \\ \\ \dot{\overline
p}&=&-\hs\kappa\hs{\overline x}-\hs(A-B)\hs{\overline p}\hs . \\
\end{array}\right\} \eqno (3.4) $$ If we eliminate $\overline p$
from this set of equations, we can obtain the mean coordinate
equation $$ \ddot{\overline x}+2\hs(\hh A-B\hh)\hs\dot{\overline
x}+\Bigl(\hh\omega^{\h 2}+(\hh A-B\hh)^{\hh 2} \Bigr)\hs{\overline
x}=0\hs . \eqno (3.5) $$ The above equation (3.5) provides the
following solution: $$ \overline x(t)=(\hh C_1\hh\cos\hh\omega\hs t+
C_2\hh\sin\hh\omega\hs t\hh)\hs e^{\hh -\hh(\hh A-B\hh)\hh t}\hs ,
\eqno (3.6) $$ where $C_1$ and $C_2$ are arbitrary constants.

\vskip 5mm

\centerline{\bf 4. Mean oscillator energy } \vskip 2mm

\par Now, we will find the time derivative from $\overline{a^{\hh
+}a}$. By applying the above equality (3.2) we will produce the
following derivative from equation (1.2): $$ \dot{\overline{a^{\hh
+}a}}+2\hs(\hh A-B\hh)\hs\overline{a^{\hh +}a}=2\hs B\hs . \eqno
(4.1) $$ We can define harmonic oscillator time-varying energy
effects inserting the following formula in equation (4.1): $$
\overline{a^{\hh +}a}=\frac{\overline
H}{\hbar\hs\omega}-\frac{1}{2}\hs . $$ Thus, the following
differential equation is derived: $$ \dot{\overline H}+2\hs(\hh
A-B\hh)\hs\overline H=\hbar\hs\omega\hs (\hh A+B\hh)\hs . \eqno
(4.2) $$ The solution is given by the equation as follows: $$
\overline H(t)=C\hs e^{\hh -\hs 2\hs(\hh A-B\hh)\hs t}+
\frac{\hh\hbar\hs\omega\hh}{2}\hs\frac{\hh A+B\hh}{\hh A-B\hh}\hs ,
\eqno (4.3) $$ where $C$ is an arbitrary constant.

\par The equation (4.2) has specific stationary solution: $$
\overline H=\frac{\hbar\hs\omega}{2}\hs\frac{\hh A+B\hh}{A-B} \hs .
\eqno (4.4) $$ Since constants $A$ and $B$ are related (2.7), the
stationary solution obeys the formula as follows: $$  \overline
H=\frac{\hbar\hs\omega}{2}\hs \frac{\hh
e^{\hh\beta\hh\hbar\hs\omega}+1\hh}
{e^{\hh\beta\hh\hbar\hs\omega}-1}=
\frac{\hh\hbar\hs\omega\hh}{2}\hs\cth\frac{\hh\beta\hs\hbar\hs\omega\hh}{2}
\hs . \eqno (4.5) $$ If it is assumed that $T$ tents to zero, than
$\overline H=\hbar\hs\omega/\hh 2$. If it assumed that $T$ increases
to infinity, than $\overline H=k\hs T$.

\vskip 5 mm

\centerline{\bf 5. Kinetic equation expressed in terms of coordinate
and momentum operators } \vskip 2mm

\par We well express the equation (1.2) in terms of operators $\hat
x$  and $\hat p$. For this purpose, we will firstly write the
equation (1.2) as follows: $$ i\hs\hbar\hskip 1.1mm\dot{\hskip
-0.5mm\hat\varrho}= \widehat
 H\hs\hat\varrho-\hat\varrho\hs\widehat H+i\hs\hbar\hs A\hs\bigl(\hh
 2\hs\hat a\hh\hat\varrho\hh\hat a^{\hh +}-\hat a^{\hh +}\hh\hat
 a\hh\hat\varrho-\hat\varrho\hh\hat a^{\hh +}\hh\hat a\hh\bigr) +
 i\hs\hbar\hs B\hs\bigl(\hh 2\hs\hat a^{\hh +}\hh\hat\varrho\hh\hat
 a-\hat a\hh\hat a^{\hh +} \hh\hat\varrho-\hat\varrho\hh\hat a\hh\hat
 a^{\hh +}\bigr)\hs . \eqno (5.1) $$

\par Since the energy operator is equal to: $$ \widehat
H=\frac{\hh\hat p^{\hh 2}}{\hh 2\hs m\hs}+ \frac{\hh\kappa\hs\hat
x^{\hh 2}}{2}\hs , \eqno (5.2)$$ we will insert it in equation (5.1)
along with formula (1.4) to obtain the following one: $$
i\hs\hbar\hs\dot{\hat\varrho}=\biggl(\frac{\hh\hat p^{\hh 2}}{\hh
2\hs m\hs}+ \frac{\hh\kappa\hs\hat x^{\hh
2}}{2}\biggr)\hs\hat\varrho- \hat\varrho\hs\biggl(\frac{\hh\hat
p^{\hh 2}}{\hh 2\hs m\hs}+ \frac{\hh\kappa\hs\hat x^{\hh
2}}{2}\biggr)\hs - $$ $$ -\hs\frac{\hh i\hs(\hh
A+B\hh)\hh}{2\hs\omega}\hs\biggl(\frac{1}{\hh m\hh} \bigl(\hh\hat
p^{\hh 2}\hs\hat\varrho-2\hs\hat p\hs\hat\varrho\hs\hat p+
\hat\varrho\hs\hat p^{\hh 2}\hh\bigr)+ \kappa\hs\bigl(\hh\hat x^{\hh
2}\hs\hat\varrho-2\hs\hat x\hs\hat\varrho\hs\hat x+
\hat\varrho\hs\hat x^{\hh 2}\hh\bigr)\biggr)\hs - $$ $$ -\hs(\hh
A-B\hh)\hs(\hh\hat x\hs\hat\varrho\hs\hat p- \hat
p\hs\hat\varrho\hs\hat x+i\hs\hbar\hs\hat\varrho\hh)\hs . \eqno
(5.3) $$

\vskip 5mm

\centerline{\bf 6. Coordinate representation }\vskip 2mm

\par As referred to coordinate function, the statistical matrix is
represented by formula $\varrho(\hh t,\hs x,\hs x^{\hh\pr}\hh)$. In
this case, there shall be produced the following coordinate and
momentum operators: $$ \hat x=x\hs ,  \hskip 15 mm \hat p=-\hs
i\hs\hbar\hs\partial_x\hs . $$ Using the above values we can write
equation (5.3) by the formula as follows: $$ \partial_{\hh
t}\hh\varrho=\frac{i\hs\hbar}{\hh 2\hs m\hh}\hs (\hh\partial_x^{\hh
2}-\partial_{x^{\pr}}^{\hh 2})\hs\varrho- \frac{i\hs\kappa}{\hh
2\hs\hbar\hh}\hs\Bigl(\hh x^2-{x^\pr}^2\Bigr) \hs\varrho+ \frac{\hh
A+B\hh}{2\hs\hbar\hs\omega}\hs\biggl(\frac{\hh\hbar^2}{m}
\hs(\hh\partial_x+\partial_{x^\pr})^2-\kappa\hs(\hh x-x^\pr)^2
\biggr)\hs\varrho\hs + \hs $$ $$ + \hs(\hh A-B\hh)\hs(\hh
1+x\hs\partial_{\hh x^{\h\pr}}+ x^{\h\pr}\hh\partial_x)\hs
\varrho\hs . \eqno (6.1) $$

\par As concerns statistical matrix physical interpretation, the
following formula $$ w(t,\hs x)=\varrho(\hh t,\hs x,\hs x\hh) \eqno
(6.2) $$ is applied for getting specific probability coefficient.

\par Now, we will add new variables $$ x_1=\frac{1}{\hh 2\hh}\hs(\hh
x+x^\pr)\hs , \hskip 15 mm x_2=x-x^\pr\hs . \eqno (6.3) $$ In this
case $$ \partial_x=\frac{1}{\hh 2\hh}\hs\partial_1+\partial_{\h
2}\hs , \hskip 15 mm \partial_{x^{\h\pr}}=\frac{1}{\hh
2\hh}\hs\partial_1-\partial_{\h 2}\hs . $$

\par Referring to statistical matrix $\varrho(\hh t,\hs x_1,\hs
x_2\hh)$ and using the above new variables we will gain the equation
as follows: $$ \partial_{\hh t}\hh\varrho=\frac{\hh
i\hs\hbar\hh}{m}\hs \partial_{\hh 1}\hh\partial_{\h 2}\hh\varrho-
\frac{\hh i\hs\kappa\hh}{\hbar}\hs x_{\h 1}\hh x_{\h 2}\hh\varrho +
\frac{\hh A+B\hh}{\hbar\hs\omega}\hs\biggl( \frac{\hbar^{\h 2}}{\hh
2\hs m\hh}\hs\partial_1^{\h 2}\hh\varrho- \frac{\hh\kappa\hh}{2}\hs
x_{\h 2}^{\h 2}\hh\varrho+ \varepsilon\hs(\hh 1+x_1\hh\partial_{\h
1}-x_2\hh\partial_{\h 2})\hs \varrho\hh\biggr)\hs , \eqno (6.4) $$
where $$ \varepsilon=\hbar\hs\omega\hs\frac{\hh A-B\hh}{A+B}=
\hbar\hs\omega\hs\th\hh\frac{\hh\beta\hs\hbar\hs\omega\hh}{2}\hs .
\eqno (6.5) $$ In this case $$ \varrho(\hh t,\hs x,\hs 0\hh)=w(t,\hs
x)\hs . \eqno (6.6) $$

\par We will try solution of equation (6.4) as follows: $$
\varrho(\hh t,\hs x_1,\hs x_2\hh)=\frac{1}{\hh 2\hs\pi\hh}\hs\int
f(\hh t,\hs k,\hs x_2\hh)\hs e^{\hh i\hh k\hh x_1}\hs dk\hs . \eqno
(6.7) $$ Reciprocal transformation $$ f(\hh t,\hs k,\hs
x_2\hh)=\int\varrho(\hh t,\hs x_1,\hs x_2\hh)\hs e^{\hh -\hh i\hh
k\hh x_1}\hs dx_1\hs . \eqno (6.8) $$ With provision for equation
(6.6) we will obtain: $$ f(\hh t,\hs 0,\hs 0\hh)=\int\varrho(\hh
t,\hs x,\hs 0\hh)\hs dx=\int w(\hh t,\hs x\hh)\hs dx=1\hs . \eqno
(6.9) $$

\par Thus, in view of function (6.8) the following equation is
formed: $$ \partial_{\hh t}\hh f=-\hs\frac{\hh\hbar\hh\hs
k\hh}{m}\hs\partial_x\h f+ \frac{\hh\kappa\hh}{\hbar}\hs
x\hs\partial_{\h k}\h f- \frac{\hh A+B\hh}{\hbar\hs\omega}\hs\biggl(
\frac{\hh\hbar^{\h 2}\hh k^2\hh}{2\hs m}+ \frac{\hh\kappa\hs x^{\h
2}}{2}+\varepsilon\hs (\hh k\hs\partial_{\h
k}+x\hs\partial_x\hh)\biggr)\hh f\hs . \eqno (6.10) $$ This equation
has an equilibrium solution which satisfies the both formulas as
follows: $$ -\hs\frac{\hh\hbar\hh\hs k\hh}{m}\hs\partial_x\h f+
\frac{\hh\kappa\hh}{\hbar}\hs x\hs\partial_{\h k}\h f=0\hs , \eqno
(6.11) $$ $$ \frac{\hh\hbar^{\h 2}\hh k^2\hh}{2\hs m}\hs f+
\frac{\hh\kappa\hs x^{\h 2}}{2}\hs f+\varepsilon\hs (\hh
k\hs\partial_{\h k}+x\hs\partial_x\hh)\hh f=0\hs . \eqno (6.12) $$

\par We will write the performance equation of the above formula
(6.11): $$ -\hs\frac{\hh m\hs dx\hh}{\hbar^{\h 2}\hh k}=
\frac{dk}{\hh\kappa\hs x\hh}\hs . $$ This equation has the solution
as follows: $$ \frac{\hh\hbar^{\h 2}\hh k^2\hh}{2\hs m}+
\frac{\hh\kappa\hs x^{\h 2}\hh}{2}=const\hs . $$ This formula
implies that the general solution of equation (6.11) takes the form
as follows: $$ f=f(E)\hs , $$ where $$ E=\frac{\hh\hbar^{\h 2}\hh
k^2\hh}{2\hs m}+ \frac{\hh\kappa\hs x^{\h 2}\hh}{2}\hs . $$ Now, we
will insert this function in equation (6.12) to gain the following
formula: $$ \frac{d\hh f}{dE}+\frac{f}{\hh 2\hs\varepsilon\hh}=0\hs
. $$ With provision for condition (6.9) this equation has the
following solution: $$ f(E)=\exp\biggl(-\hs\frac{E}{\hh
2\hs\varepsilon\hh}\biggr)\hs . $$ Thus, the equilibrium solution of
equation (6.10) takes the form as follows: $$ f(\hh k,\hs
x\hh)=\exp\Biggl(-\hs\frac{1}{\hh 2\hs\varepsilon\hh}\hs
\biggl(\frac{\hh\hbar^{\h 2}\hh k^2\hh}{2\hs m}+ \frac{\hh\kappa\hs
x^{\h 2}\hh}{2}\biggr)\Biggr)\hs . \eqno (6.13) $$

\par We will find the equilibrium statistical matrix by formula
(6.7) to obtain the following equation: $$ \varrho(\hh x_1,\hs
x_2\hh)=\frac{1}{\hh 2\hs\pi\hh}\hs\int \exp\Biggl(-\hs\frac{1}{\hh
2\hs\varepsilon\hh}\hs \biggl(\frac{\hh\hbar^{\h 2}\hh k^2\hh}{2\hs
m}+ \frac{\hh\kappa\hs x_{\hh 2}^{\h 2}\hh}{2}\biggr)\Biggr) \hs
e^{\hh i\hh k\hh x_1}\hs dk\hs . $$ On integrating the following
formula is produced: $$ \varrho(\hh x_1,\hs
x_2\hh)=\sqrt{\frac{\hh\alpha\hh}{\pi}}\hs \exp\biggl(\hh
-\hs\alpha\hs x_{\hh 1}^{\h 2}- \frac{\hh\sigma^{\h 2}\hh x_{\hh
2}^{\h 2}}{4\hs\alpha}\biggr)\hs , \eqno (6.14) $$ where $$
\alpha=\sigma\hs\th\frac{\hh\beta\hh\hbar\hs\omega\hh}{2}\hs ,
\hskip 15 mm \sigma=\frac{\hh m\hs\hs\omega\hh}{\hbar}\hs . $$

\par Using formulas (6.3) we will get the equation as follows: $$
\varrho(\hh x,\hs x^{\h\pr}\hh)=\sqrt{\frac{\hh\alpha\hh}{\pi}}\hs
\exp\biggl(\hh -\hs\frac{\hh\alpha\hs(\hh x+x^{\h\pr})^{\h 2}}{4}-
\frac{\hh\sigma^{\h 2}\hs (\hh x-x^{\h\pr})^{\h
2}}{4\hs\alpha}\biggr)\hs . \eqno (6.15) $$

\par Using formula (6.2) we will get respective equilibrium
probability coefficient [17] $$ w(\hh
x\hh)=\sqrt{\frac{\hh\alpha\hh}{\pi}}\hs \exp\Bigl(\hh -\hs\alpha\hs
x^{\h 2}\hs\Bigr)\hs . \eqno (6.16) $$

\vskip 5mm

\centerline{\bf 7. Momentum representation } \vskip 2mm

\par As referred to momentum function, the coordinate and momentum
operators are represented by the formulas as follows: $$ \hat
x=i\hs\hbar\hs\partial_{\h p}\hs , \hskip 15 mm \hat p=p\hs . $$ In
view of momentum representation, the statistical matrix is
formulated as $\varrho(\hh t,\hs p,\hs p^{\hh\pr}\hh)$. This form
may be applied for obtaining equation (5.3) as follows: $$
\partial_{\hh t}\hh\varrho=-\hs\frac{i}{\hh 2\hs\hbar\hs m\hh}\hs
\Bigl(\hh p^{\hh 2}-{p^{\hh\pr}}^{\h 2}\Bigr)\hs\varrho+
\frac{i\hs\hbar\hs\kappa}{\hh 2\hh}\hs\Bigl(\hh\partial_{\h p}^{\hh
2}- \partial_{\h p^{\h\pr}}^{\hh 2}\Bigr)\hs\varrho- \frac{\hh
A+B\hh}{2\hs\hbar\hs\omega}\hs\biggl(\frac{\hh 1\hh}{m} \hs(\hh
p-p^{\h\pr})^2-\kappa\hs\hbar^{\hh 2}\hs(\hh\partial_{\h p}+
\partial_{\h p^{\h\pr}})^2\biggr)\hs\varrho\hs - \hs $$ $$ -\hs(\hh
A-B\hh)\hs(\hh 1+p\hs\partial_{\hh p^{\h\pr}}+
p^{\h\pr}\hh\partial_p)\hs \varrho\hs . \eqno (7.1) $$

\par As concerns physical interpretation of statistical matrix
$\varrho(\hh t,\hs p,\hs p^{\hh\pr}\hh)$, the following formula
refers to $$ w(t,\hs p)=\varrho(\hh t,\hs p,\hs p\hh) \eqno (7.2) $$
probability density applied for detecting the state when an
oscillator have impulse $p$.

\vskip 5mm

\centerline{\bf 8. Wigner function } \vskip 2mm

\par In order to better appreciate the physical significance of
various kinetic state summands we will derive the equation for
Wigner function $w=w(\hh t,\hs x,\hs p\hh)$, which is specified as a
quantum analog of classical distribution function and may be defined
by applying statistical matrix $\varrho=\varrho(\hh t,\hs x,\hs
x^{\hh\pr}\hh)$ specified by the relation as follows: $$ w(\hh t,\hs
x,\hs p\hh)=\frac{1}{\hh 2\hs\pi\hh}\hs\int \varrho\Bigl(\hh t,\hs
x+\frac{1}{\hh 2\hh}\hs\hbar\hs q,\hs x-\frac{1}{\hh
2\hh}\hs\hbar\hs q\hh\Bigr)\hs e^{\hh -\hh i\hh p\hh q\hh}\hs dq\hs
. \eqno (8.1) $$ If the statistical matrix depends on $x_1$ and
$x_2$, than $$ w(\hh t,\hs x,\hs p\hh)=\frac{1}{\hh
2\hs\pi\hs\hbar\hh}\hs\int \varrho(\hh t,\hs x_1=x,\hs x_2\hh)\hs
e^{\hh -\hh i\hh p\hh x_2\hh/\hh\hbar}\hs dx_2\hs . \eqno (8.2) $$

\par Reciprocal transformation $$ \varrho(\hh t,\hs x_1,\hs
x_2\hh)=\int w(\hh t,\hs x_1,\hs p\hh)\hs e^{\hh i\hh p\hh
x_2\hh/\hh\hbar}\hs dp\hs . \eqno (8.3) $$ Since there is formula
(6.9) $$ \int\varrho(\hh t,\hs x_1,\hs 0\hh)\hs dx_1=1\hs , $$
Wigner function satisfies specific normalization requirement $$ \int
w(\hh t,\hs x,\hs p\hh)\hs dx\hs dp\hs=1\hs . \eqno (8.4) $$

\par For the purpose of Wigner function, we will derive the
following formula from equation (6.4) formulated for statistical
matrix $\varrho(\hh t,\hs x_1,\hs x_2\hh)$ $$
\partial_tw=-\hs\frac{\hh p\hh}{m}\hs\partial_x\h w+ \kappa\hs
x\hs\partial_p\h w+ \frac{\hh A+B\hh}{2\hs\hbar\hs\omega}\hs\biggl(
\frac{\hbar^{\hh 2}}{2\hs m}\hs\partial_{\h x}^{\hh 2}\hh w+
\frac{\hh\hbar^{\hh 2}\hh\kappa\hh}{2}\hs\partial_{\h p}^{\hh 2}\hh
w+ \varepsilon\hs(\hh 2+x\hs\partial_x+p\hs\partial_p\hh)\hs
w\biggr)\hs . \eqno (8.5) $$ The equation produced is much different
from its quantum analog of Fokker-Planck equation. Summands
containing derivatives $\partial_{\h x}^{\hh 2}\hh w$ and
$\partial_{\h p}^{\hh 2}\hh w$ may be interpreted as those
describing phase space diffusion. And still, it is necessary to add
that it is rather hard to appreciate physical significance of the
formula in parentheses that follows coefficient $\varepsilon$.

\par Equilibrium solution of equation (8.5) should at the same time
refer to the following equations: $$ -\hs\frac{\hh
p\hh}{m}\hs\partial_x\h w+ \kappa\hs x\hs\partial_p\h w=0\hs , \eqno
(8.6) $$ $$ \frac{\hbar^{\hh 2}}{2\hs m}\hs\partial_{\h x}^{\hh
2}\hh w+ \frac{\hh\hbar^{\hh 2}\hh\kappa\hh}{2}\hs\partial_{\h
p}^{\hh 2}\hh w+ \varepsilon\hs(\hh
2+x\hs\partial_x+p\hs\partial_p\hh)\hs w=0\hs . \eqno (8.7) $$
General solution of the equation (8.6) derives by the function as
follows: $$ w=w(E)\hs , \eqno (8.8) $$ where $$ E=\frac{p^{\hh
2}}{\hh 2\hs m\hh}+\frac{\hh\kappa\hs x^{\hh 2}}{2}\hs . $$ We will
insert the above function in equation (8.7). Thus, we will get the
following equation: $$ E\hs\frac{d^{\hh 2}w}{dE^{\hh 2}}+ (\hh
1+\mu\hs E\hh)\hs\frac{dw}{dE}+\mu\hs w=0\hs , \eqno (8.9) $$ where
$$ \mu=\frac{2}{\hh\hbar\hs\omega\hh}\hs\th\hs
\frac{\hh\beta\hs\hbar\hs\omega\hh}{2}\hs . \eqno (8.10) $$ Solution
of this equation derives the function as follows: $$ w(E)=C\hs
e^{\hh -\hh\mu\hh E}\hs . \eqno (8.11) $$

\par Wigner function may be derived by applying formula (8.2),
providing that specific equilibrium function (6.14) is inserted in.
In view of this we shall obtain the following equation: $$ w(\hh
x,\hs p\hh)=\frac{1}{\hh 2\hs\pi\hs\hbar\hh}\hs
\sqrt{\frac{\hh\alpha\hh}{\pi}}\hs\int \exp\biggl(\hh -\hs\alpha\hs
x^{\h 2}- \frac{\hh\sigma^{\h 2}\hh x_{\hh 2}^{\h
2}}{4\hs\alpha}\biggr)\hs \hs e^{\hh -\hh i\hh p\hh
x_2\hh/\hh\hbar}\hs dx_2\hs . \eqno (8.12) $$ On integrating we will
get the equilibrium function $$ w(\hh x,\hs
p\hh)=\frac{\hh\mu\hs\omega\hh}{2\hs\pi}\hs
\exp\Biggl(-\hs\mu\hs\biggl(\frac{p^{\hh 2}}{\hh 2\hs m\hh}+
\frac{\hh\kappa\hs x^{\hh 2}}{2}\biggr)\Biggr)\hs . \eqno(8.13) $$
This function may be formulated by the equation as follows: $$ w(\hh
x,\hs p\hh)= \sqrt{\frac{\hh\alpha\hh}{\pi}}\hs \exp\Bigl(\hh
-\hs\alpha\hs x^{\h 2}\hs\Bigr)\hs
\sqrt{\frac{\hh\gamma\hh}{\pi}}\hs \exp\Bigl(\hh -\hs\gamma\hs p^{\h
2}\hs\Bigr)\hs , \eqno (8.14) $$ where $$ \gamma=\frac{\mu}{\hh 2\hs
m\hh}\hs . \eqno (8.15) $$

\par Then, we will find the mean value $$ \overline{\hh x^{\hh
2}}\hs\hs\overline{\hh p^{\hh 2}}= \int x^{\hh 2}\hs p^{\hh 2}\hs
w(\hh x,\hs p\hh)\hs dx\hs dp\hs . \eqno (8.16) $$ The above
computation gives the following formula $$ \overline{\hh x^{\hh
2}}\hs\hs\overline{\hh p^{\hh 2}}=\frac {\hh\hbar^{\hh
2}}{4}\hs\biggl(\frac{\hh e^{\hh\beta\hh\hbar\hh\omega}+1\hh} {\hh
e^{\hh\beta\hh\hbar\hh\omega}-1\hh}\biggr)^2\hs . \eqno (8.17) $$
This formula has the following result. If $T=0$, the uncertainty is
equal to $\overline{\hh x^{\hh 2}}\hs\hs\overline{\hh p^{\hh 2}}=
\hbar^{\hh 2}/4$. If $T\to\infty$, than $\overline{\hh x^{\hh
2}}\hs\hs\overline{\hh p^{\hh 2}}= m\hs (\hh k\hs T\hh)^{\hh
2}/\hh\kappa$.

\vskip 5mm

\centerline{\bf 9. Conclusion } \vskip 2mm

\par The equation proposed by Lindblad for the purpose of the
statistical operator describing nonequilibrium state of quantum
harmonic oscillator is hereby considered. Initially, the statistical
matrix equation in energy representation and diagonal matrix element
equation have been derived from the equation concerned. Specific
formulas appreciating physical significance of Lindblad equation
coefficients have been formulated. Then, the mean coordinate
equation has been derived to find any general solution. It was
demonstrated that the mean coordinate exponentially decays in time.
The mean oscillator energy equation has been derived to obtain the
general solution and mean equilibrium energy value has been found.
Lindblad equation has been formulated by applying coordinate and
momentum operators. The coordinate representation statistical matrix
equation has been obtained. The equilibrium statistical matrix
formula has been derived from this equation. The momentum
representation statistical matrix equation has been formulated.
Wigner function equation has been obtained and the respective
equilibrium state function has been found. Various temperature
uncertainty relations have been found by applying Wigner equilibrium
function.\vskip 5mm

\centerline{\bf References }

\vskip 2mm

\noindent [1] John von Neumann, Mathematical Basis of Quantum
Mechanics, Nauka, M, 1964. \vskip 1mm

\noindent [2] D.I.Blokhintzev, Fundamental Quantum Mechanics, Higher
School, M, 1961. \vskip 1mm

\noindent [3] L.D.Landau, E.M.Lifshits, Quantum Mechanics, Nauka, M,
1963. \vskip 1mm

\noindent [4] K.Blum, Density Matrix Theory and Application, Plenum,
New York, London, 1981. \vskip 1mm

\noindent [5] R.L.Stratonovich, Nonlinear Nonequilibrium
Thermodynamics, Nauka, M, 1985. \vskip 1mm

\noindent [6] Y.R.Shen, Phys. Rev. 155 (1967) 921-931. \vskip 1mm

\noindent [7] M.Grover, R.Silbey, Chem. Phys. 52:3 (1970) 2099-2114,
54:1 (1971) 4843-4857. \vskip 1mm

\noindent [8] A.Kossakowski, Rep. Math. Phys. 3:4 (1972) 247-274.
\vskip 1mm

\noindent [9] V.Gorini, A.Kossakowski and E.C.G.Sudarshan, J. Math.
Phys. 17:5 (1976) 821-826. \vskip 1mm

\noindent [10] G.Lindblad, Commun. Math. Phys. 48:2 (1976) 119-130.
\vskip 1mm

\noindent [11] V.Gorini, A.Frigerio, N.Verri, A.Kossakowski and
E.C.G.Sudarshan, Rep. Math. Phys. 13:2 (1978)\vskip 1mm 149-173.
\vskip 1mm

\noindent [12] R.Alicki, K.Lendy, Quantum Dynamical Semigroups and
Applications. Lecture Notes in Physics. \vskip 1mm vol. 286,
Springer-Verlag, Berlin, 1987. \vskip 1mm

\noindent [13] J.L.Neto, Ann. Phys. (NY) 173:2 (1987) 443-461.
\vskip 1mm

\noindent [14] B.V.Bondarev, Physica A 176:2 (1991) 366-386. \vskip
1mm

\noindent [15] B.V.Bondarev, Physica A 183:1-2 (1992) 159-174.
\vskip 1mm

\noindent [16] B.V.Bondarev, Teoret. Mat. Fis. 100:1 (1994) 33-43.
\vskip 1mm

\noindent [17] L.D.Landau, E.M.Lifshits, Statistical Physics, Nauka,
M, 1964.

\end{document}